\documentclass[pdflatex,sn-basic]{sn-jnl}

\usepackage{soul}
\usepackage{hyperref}
\usepackage{mathtools} 
\usepackage{siunitx} 
\usepackage{dsfont}
\usepackage{graphicx} 
\usepackage{amsmath}
\usepackage{tabularx}
\usepackage[algo2e,ruled,vlined,linesnumbered]{algorithm2e}

\newcommand{\sC}{\mathcal{C}}
\newcommand{\sE}{\mathcal{E}}

\newcommand{\sQ}{\mathcal{Q}}

\newcommand{\expt}{Exp3 }
\newcommand{\expte}{{TEF }}

\mathchardef\hyph="2D

\jyear{2023}%

\theoremstyle{thmstyleone}%
\newtheorem{theorem}{Theorem}
\theoremstyle{thmstyletwo}%
\newtheorem{remark}{Remark}%

\theoremstyle{thmstylethree}%

\begin{document}
\title[Improving Sequential Query Recommendation with Immediate User Feedback]{Improving Sequential Query Recommendation with Immediate User Feedback}
\author*[1]{\fnm{Shameem~A} \sur{Puthiya~Parambath}}\email{sham.puthiya@glasgow.ac.uk}
\author[1]{\fnm{Christos} \sur{Anagnostopoulos}}\email{christos.anagnostopoulos@glasgow.ac.uk}
\author[1]{\fnm{Roderick} \sur{Murray-Smith}}\email{roderick.murray-smith@glasgow.ac.uk}
\affil[1]{\orgdiv{School of Computing Science}, \orgname{University of Glasgow}, \orgaddress{\country{UK}}}

\abstract{
We study the problem of predicting the next query to be recommended in interactive data exploratory analysis to guide users to correct content.
Current query prediction approaches are based on sequence-to-sequence learning, exploiting past interaction data.
However, due to the resource-hungry training process, such approaches fail to adapt to immediate user feedback.
Immediate feedback is essential and considered as a signal of the user's intent.
We contribute with a novel query prediction ensemble mechanism, which adapts to immediate feedback relying on Multi-Armed Bandits (MAB) framework. Our mechanism, an extension to the popular \expt algorithm, augments Transformer-based language models for query predictions by combining predictions from experts, thus dynamically building a candidate set during exploration. Immediate feedback is leveraged to choose the appropriate prediction in a probabilistic fashion. We provide comprehensive large-scale experimental and comparative assessment using a popular online literature discovery service, which showcases that our mechanism (i) improves the per-round regret substantially against state-of-the-art Transformer-based models and (ii) shows the superiority of causal language modelling over masked language modelling for query recommendations.
}

\maketitle
\section{Introduction}
The data exploration process consists of incremental knowledge extraction and gathering where users (like data scientists and analysts) sequentially explore the data space through sequences of queries predicted and suggested by online algorithms. It is evidenced that query recommendation speedups the data exploration process by addressing the need for disambiguation of queries and directing analysts to acquire the intended knowledge \citep{DehghaniRAF17}. 
Moreover, query prediction supports the management of distributed data systems in resource-efficient decision-making \citep{parambath2021maxutility}.
As pointed out in \citet{parambath2021maxutility}, an online interactive data exploration session starts with a user issuing an initial query pertaining to a specific topic. Then, the system responds with relevant results along with a \textit{recommendation of a suitable query} to be executed in the next round of the exploration process.
In such online interaction, the system can capture the immediate user feedback to the recommended query, e.g., the user accepts the recommended query in the form of a `click'. This feedback serves as a signal of the user's intent; positive feedback (acceptance by a click) signals that the recommended query aligns with the user's intent, while negative feedback (no click) signals the opposite. Typical application scenarios involve literature discovery on the web, among others \cite{parambath2021maxutility}.

In the online learning paradigm, the user feedback to the recommended action is considered as the \textit{reward} received by the online algorithm \cite{lattimore2020bandit,cesa2006prediction}. At each step, the algorithm recommends an action, i.e., a query in our case, and a click on the recommended query indicates a reward for a correct prediction, thus, correct inference of the user's intent. The online algorithm can make use of each reward received at each step to improve the query recommendation by gradually adapting its recommendation strategy. The overarching aim of this process is to maximize the cumulative reward captured by predictions (or, equivalently, minimize the cumulative loss) over a finite period of time.

The above-mentioned online user-system interactive learning process 
can be modelled through the Multi-Armed Bandits (MAB) framework with countably many arms (bandits) \citep{KalvitZ20,bayati_2020}. 
A MAB framework models the trade-off between exploration and exploitation over a sequence of actions (queries in our case) \cite{lattimore2020bandit,cesa2006prediction}. The MAB framework provides the principles of handling the inherent uncertainty about the suitable query to be predicted at each step of the exploratory data analysis and knowledge extraction process.

\subsection{The Query Prediction Problem in Multi-arm Bandits Context}
Let us first discuss the application of a standard stochastic MAB framework to the online query prediction problem. 
At each discrete step $t= 1,\dots,T$, depending on the currently executing query, the (MAB-based) system chooses a query $q_{t}$ to be executed from a very large set of historical queries $\sQ$. All the past executed queries until the current query in a session is referred to as \textit{query context}, i.e., the sequence $(q_{1}, q_{2}, \ldots, q_{t})$. The query context indicates the user's intent. Specifically, each predicted query $q_{t}$ is associated with an intrinsic utility, which depends on the user's intent captured by the current query context. Depending on the user's intent, the user can either accept or ignore (reject) the recommendation of the predicted query, thus, revealing a reward $r_{t}$ to the system at the $t$-th step. 
In this fashion, the system repeatedly predicts the next query to be recommended, observes the collected rewards by the user so far, and tries to maximize the expected cumulative rewards over time. In the standard stochastic MAB, there are $m > 0$ fixed unknown reward probability distributions associated with $m$ queries available in the query set, i.e., $m = \lvert \sQ \rvert$. 
The distributions are assumed to be independent and identical. The objective is to identify the \textit{optimal} query with the highest expected reward distribution. This indicates that the MAB algorithm should explore the whole \textit{query space} to find the optimal query with the highest reward value. 

However, in practice, it is unrealistic to assume that a user progresses with a fixed number of queries during the data exploration process regardless of the past queries issued in the session. Moreover, the predicted query to be recommended should depend on the sequence of the past issued queries so far, reflecting a continuation and progression context of the exploratory data analysis, which this context is bound to change in every round. In fact, depending on the knowledge extraction and gathering process needs, a user might be interested in queries pertaining to different aspects of a topic of interest. For instance, consider a user who is exploring the research topic of `multi-armed bandits'. The user could start the session with a simple query and dive into diverse subtopics of MAB, like applications, different MAB algorithms, reward distributions, action structures, etc., as the knowledge-gathering process progresses. Hence, by repeatedly recommending the next query pertaining to a single subtopic, disregarding the sequence(s) of past issued queries, is not expected to achieve positive responses (acceptances) from the user. 
Evidently, this hinders user satisfaction and system usability. Therefore, the stochastic reward assumption used in the standard MAB mechanisms does not successfully operate in online knowledge-gathering processes.

In this paper, we depart from these limitations of standard stochastic MAB mechanisms and tackle the query prediction problem in a non-stochastic fashion. Our principle is based on the autoregressive nature of the knowledge-gathering process, where we make no probabilistic assumptions on how the rewards, corresponding to different query predictions, are generated. In particular, the rewards may depend on previously predicted and executed queries, which can also be adversarial. Earlier work on query prediction using MAB assumes a fixed query set, which does not take into account the dynamics of the data exploration process \citep{parambath2021maxutility}.
Moreover, the aforementioned approach is based on the similarities between queries in the query context and the stochastic reward structure for the queries. Both of these aspects fail to model the autoregressive nature of the knowledge-gathering process. It should also be noted that standard adversarial MAB algorithms like \expt \citep{auer2002nonstochastic,cesa2006prediction} cannot be directly used in query prediction tasks. This is due to the fact that such algorithms are designed to work with a \textit{fixed and known} set of queries only. Nonetheless, query predictions in knowledge-gathering require the next/predicted query to depend on the query context, which cannot be assumed to be known beforehand.

Due to the sequential nature of knowledge-gathering, autoregressive time series models for text generation have demonstrated promising results in many practical knowledge extraction tasks as shown in  \citet{DehghaniRAF17,ren2018conversational,ahmad2019context,jiang2018rin,wu2018query,mustar2020using}.
Autoregressive query prediction models take into account the previously issued queries (in the same session) when proceeding with the 
query prediction to filter out irrelevant queries. This helps the user with the proper exploration of the data space.
The state-of-the-art query prediction algorithms make use of the deep Recurrent Neural Networks (RNN) based autoregressive models to extract certain linguistic patterns from query logs and combine them with historical interaction data to predict the next query. Unfortunately, standard RNN models like Long-Short Term Memory (LSTM) are designed for sequential data processing and are limited by the relatively short span of the query dependency that can be captured by such networks \citep{mustar2020using,mustar21_tis}. On the other hand, the recently proposed Transformer networks \citep{vaswani2017attention} can process data in parallel. In addition, the standard deep autoregressive models are not designed for real-time interactive query predictions and recommendations as the learning is achieved over historical data. Hence, immediate user feedback is not considered in the decision-making. All in all, a query prediction mechanism in sequential knowledge gathering should be able to \textit{adapt} to the immediate user feedback and query context to continuously improve its performance, as it will be demonstrated by our mechanism and our experiments.

\subsection{Contribution} 
We contribute with a MAB mechanism dealing with the sequential query prediction problem. Our mechanism combines the effectiveness of Transformer-based auto-regressive models and the adaptivity of the MAB expert algorithms \citep{cesa1997use,cesa2006prediction} under non-stochastic rewards.
We adopt an ensemble of state-of-the-art Transformer-based auto-regressive models for causal language modelling, hereinafter referred to as the \textit{experts}, and propose a next-query prediction algorithm that adapts to immediate user feedback. In standard non-stochastic MAB approaches \citep{auer2002nonstochastic,lattimore2020bandit}, the number of possible queries is fixed and known to the prediction mechanism in advance. On the other hand, in query prediction with Transformers, given the current query context, the number of possible candidate queries, even from a single expert, can be countably many, with varying relevance probabilities. Most importantly, possible candidate queries \textit{change} in every step depending on the query context.

We first show that predicting and recommending the top query from a single expert (e.g., an autoregressive model) is not always the best strategy (\autoref{tab:illustrated_example}). Then, departing from this fundamental outcome, our contribution is the
introduction of a non-stochastic MAB algorithm, an extension to the popular \expt algorithm, that \textit{dynamically} chooses a \textit{variable size} candidate set via an ensemble of different experts' recommendations/predictions extending the capacity of \expt by \cite{auer2002nonstochastic} toward non-fixed and unknown sizes of sets of dynamically changing queries. In Section~\ref{sec:algo}, we provide a theoretical analysis and prove the \textit{regret bound} of our MAB algorithm showing that it achieves better regret than the standard baseline \expt algorithm, and matches the best attainable regret in adversarial settings \citep{audibert2009minimax}. In Section~\ref{sec:exp}, we provide extensive experiments and comparative assessment against mechanisms and baseline models found in the literature namely: \expt \cite{auer2002nonstochastic}, GPT2 \cite{radford2019language}, Transformer-XL \cite{dai2019transformer}, CTRL \cite{keskarCTRL2019}, BERT \cite{mustar2020using}, BART \cite{lewis2020bart}, and HRED \cite{sordoni2015hierarchical}, using query logs from online literature discovery services. Furthermore, our experiments show that fine-tuning publicly available pre-trained Transformer models trained on a huge corpus provide marginally better performance than learning a model from scratch with limited query log data. Finally, Section~\ref{sec:conc} concludes the article with future research directions. (Note: in the remainder, the terms recommended query and predicted query are used interchangeably, depending on the context.)

\section{Related Work}
\label{sec:rel_work}
In this section, we review learning models adopted for query prediction providing a representative running example and elaborate on MAB models for on-line query prediction taking into account the user feedback.
\subsection{Autoregressive \& Causal Language Models for Query Prediction}
The standard unsupervised language Autoregressive (AR) models' objective is predicting the next word token (comprising a query), having processed all the previous ones. Such models have been popular in various Natural Language Processing (NLP) tasks, while recently they have been used for knowledge retrieval, query prediction, query reformulation, and document ranking \citep{DehghaniRAF17, ren2018conversational}. Specifically, Recurrent Neural Networks (RNN) are AR models adopted for query prediction due to their ability to process variable length input sequences \citep{ahmad2019context, jiang2018rin, wu2018query,sordoni2015hierarchical}.
\citet{DehghaniRAF17} augmented the RNN-based sequence-to-sequence (seq2seq) models for query prediction with a query-aware attention mechanism \citep{BahdanauCB14} and a pointer network \citep{vinyals2015pointer} to deal with the Out-of-Vocabulary words.
\citet{wu2018query} proposed a method to integrate historical user feedback in RNN-based query prediction approaches. The proposed method uses historical interaction data to create embeddings and attention scores for queries by combining query-tokens associated with search result contents (e.g., hyperlinks), and corresponding `click' streams. Depending on whether a hyperlink is `clicked' or not, both positive and negative feedback is learned for different queries. Although such an approach makes use of historical user interaction data, the prediction strategy is not online and does not adapt to immediate feedback.

The authors in \citet{ahmad2019context} introduced a context attentive document-ranking and query prediction algorithm. Such algorithm encodes the query and search session activities into search context representation via a two-level RNN. Query and document embeddings are constructed from pre-trained word embeddings using bi-directional RNN with an inner attention mechanism. The search context embeddings are obtained by passing the query and relevant documents through a LSTM model. The query prediction consists of a series of word-level predictions that forecast the next word, given the current query and search context.

Query prediction is also studied under query reformulation. \citet{ren2018conversational} studied the query reformulation problem within the context of conversational information retrieval using a LSTM with self-attention for seq2seq modelling. Similarly, \citet{jiang2018rin} framed the query prediction task as a query reformulation one. At each step $t$, the new query to be predicted is a reformulation of the past issued queries until step $(t-1)$. The query generator in \citet{jiang2018rin} is an RNN-based decoder that generates the next query based on the context vector formed by query reformulation embeddings.
Recently, with the advent of deep neural networks, many query prediction and recommendation algorithms have been proposed. The interested reader could refer to overview papers for query recommendation in \citet{DehghaniRAF17,ren2018conversational,ahmad2019context,jiang2018rin,wu2018query} and the references therein.

The parallel processing capabilities and the ability to capture complex query dependencies yield the Transformer networks \citep{vaswani2017attention} popular for Causal Language Modelling (CLM) tasks.
Fundamentally, the CLM objective is predicting the next token following a sequence of tokens such that the prediction for the current position depends \textit{solely on the known tokens} in the positions before the current position. The CLM objective is well suited for sequential query recommendations, as a Transformer network can access all the queries until the current round.
Recently, \citet{mustar2020using,mustar21_tis} proposed bidirectional Transformer networks, like BERT \citep{devlin2019bert} and BART \citep{lewis2020bart} for query predictions. The empirical study in the BERT and BART papers showed a significant improvement in performance over the standard RNN-based models when pre-trained Transformer networks are fine-tuned for query prediction.
In particular, \citet{mustar21_tis} showed that even training hierarchical Transformer models from scratch on limited data under-performs compared to pre-trained models. However, BERT and BART are trained using masked language modelling objectives, and thus, the prediction for the current position makes use of random tokens in preceding and following positions. In the context of query prediction, we show that CLM is better suited as the Transformer network cannot `look into the future'. Nevertheless, \citet{mustar21_tis} does not make use of immediate feedback when predicting the next query, thus, being unaware of the current user interactions and intention for knowledge extraction in exploratory tasks.

The current state-of-the-art AR models for query prediction recommend the top queries given a query context. Nonetheless, our preliminary analysis showed that combining different AR models, like Transformer-based models, query prediction is drastically enhanced by the query context awareness compared to employing single models. 

\textbf{Representative Example:} 
Consider the following representative example. We fed two consecutive queries: 
\emph{$q_1$ = `factors that affect getting life insurance'} and \emph{$q_2$ = `purchasing factors in getting life insurance'} from a randomly selected user session from our query logs to two different Transformer-based networks (Model 1 and Model 2) for next query prediction. 
The results obtained based on the top-2 query recommendations are provided in Table~\ref{tab:illustrated_example}.
From the query context $(q_1,q_2)$, it is evident that the user is interested in gathering knowledge regarding different aspects of getting life insurance. In that regard, both models suggest very relevant yet diverse terms as the predicted query keywords. But, as the top queries from the two models differ in the information content and direct the user to different search results, the simple strategy of recommending the top prediction result(s) from one model might not lead the user to reach the intended result. A better strategy could be to recommend from a mixture of the top results and then adapt the prediction based on the immediate user feedback (acceptance or rejection of the current forecast). 
In the example, the \textit{actual} queries following the first two queries
are \emph{$q_3$ = `factors influencing the purchase of life insurance'} and \emph{$q_4$ = `factors influencing the insurance amount'}. The second query (Top-2) from the Model 1 and the first one (Top-1) from Model 2 convey the user's correct intention.

\begin{table}[t]
\centering
\caption{Top-2 Recommendations (Representative Example).}
\label{tab:illustrated_example}
\begin{tabular}{@{}p{5.6cm} p{5.6cm}@{}}
    \toprule
    Model 1's prediction    &     Model 2's prediction  \\
    \midrule
    Top-1: \textit{factors that influence the profitability of insurance} &  Top-1: \textit{factors influencing the amount of insurance} \\
    Top-2: \textit{factors that influence purchasing decisions of insurance} & Top-2: \textit{factors that affects getting disability insurance} \\
    \bottomrule
\end{tabular}
\end{table}

In conclusion, the principle differences between our model and the state-of-the-art AR models in the context of query prediction and recommendation are: \emph{(i)} instead of simply recommending the top predicted query from a single AR model, we dynamically create a candidate set of predicted queries by combining top queries from different models. Then, we recommend a query probabilistically by on-line estimating a probability distribution over this set of candidate queries; \emph{(ii)} state-of-the-art AR models do not make use of immediate feedback, thus, being unaware of the current user intentions in the knowledge discovery process, whereas we exploit immediate user feedback to estimate and gradually adapt the probability distribution over the set of candidate queries w.r.t. feedback.

\subsection{Multi-armed Bandit Systems for Query Prediction}
MAB is a popular framework to model interactive learning applied, mostly, to web-based services like ad recommendation, dynamic pricing, digital health, etc \cite{bouneffouf2020survey,lattimore2020bandit}.
In query prediction, MAB provides a framework to develop on-line learning strategies that adapt to immediate user feedback. The user feedback is assumed to be the \textit{reward} for the recommendation of the predicted query. The MAB setting for query prediction can be envisaged as a two-player interactive game between the learner (algorithm) and the environment (e.g., user), where the following steps are repeated in a sequence of rounds $t= 1,\dots, T$:
\begin{itemize}
\item For each query $q \in \sC$, the environment chooses a loss vector;
\item The learner chooses a query $q_t \in \sC$ and recommends that to the user;
\item The learner observes the reward $r_t$ for the chosen query $q_t$ and adapts its learning method.
\end{itemize}
As mentioned, the reward is observed as click/acceptance. For the correct prediction, the user clicks on the recommended query (accept) and ignores it otherwise (reject). The learner's goal is to predict queries so that their total reward is as high as possible.

The standard MAB algorithms assume that the number of queries is fixed and relatively small compared to the number of steps. However, models like Transformers can generate a relatively large number of predicted queries that can be used as possible candidates for recommendation. Hence, in this context, the candidate queries in a MAB algorithm can be countably many for query prediction and recommendation.
In countably many candidate queries (corresponding to armed bandits in MAB terminology), the MAB algorithm has to deal with the discovery-exploitation trade-off during the exploration phase \citep{carpentier2015simple}, apart from the conventional exploration-exploitation trade-off. 
Several MAB algorithms for countably many or infinite armed bandits have been proposed. 
Fundamentally, \citet{berry1997bandit} proposed $k\hyph failure$ and $m\hyph run$ strategies. 
In a $k\hyph failure$ strategy, an arm is played until it incurs $k$ failures. 
In $m\hyph run$ strategy, $1\hyph failure$ is used until $m$ arms are played or $m$ successes are obtained. Another 
strategy to deal with infinite arms is \textit{pre-selection}, where a subset of $k$ arms is selected, and then standard MAB algorithms can be applied. \citet{wang2008algorithms} proposed to select $k$ randomly chosen arms for exploration and exploitation as a function of the current step, while \citet{yinglun2020} proposed to choose $k$ arms uniformly at random.
The authors in \citet{bayati_2020} proposed a greedy algorithm that selects arms uniformly at random without replacement. Recently, \citet{parambath2021maxutility} proposed a stochastic MAB algorithm for query prediction and recommendation by selecting a candidate set using the utility of the queries. The work in \citet{parambath2021maxutility} is fundamentally different from ours as their approach deals with a pre-selection strategy to choose the queries to run standard stochastic MAB algorithms. 
Nonetheless, the approach in \citet{parambath2021maxutility} does not consider the AR nature of the query process, as elaborated earlier. The reader could refer to \citet{KalvitZ20,yinglun2020,bayati_2020,kleinberg_2019,lattimore2020bandit,parambath2021maxutility} and references therein for a comprehensive review on MAB algorithms.

Finally, we report on the \textit{sleeping bandits} approach to provide a spherical review of the literature. Such an approach is closely related to the query prediction problem, albeit with significant differences. The standard adversarial sleeping bandit is studied from the sleeping experts’ point of view and not from the sleeping action point of view \citep{kleinberg2010regret,kanade2009sleeping,saha2020improved}. Specifically, in our context, experts are Transformer-based oracles, where the actions are the recommended queries from the oracles at any step. The oracles themselves will not be unavailable (\textit{sleeping}) at any point in time. 
Moreover, the available queries for different oracles at any point in time can be different or the same. Once the queries become available, they will be added to the candidate set and will never become unavailable again. The fundamental difference is, then, that none of the sleeping bandits' works considers countably many action sets of arms; thus, they cannot be adopted in our context.

Overall, the main difference between our approach and the standard adversarial MAB approaches is that, in the standard adversarial settings, the number of queries is assumed to be finite and known beforehand. 
In our problem, the number of queries is \textit{dynamic}, it can \textit{change} in every round, and it is \textit{not known} beforehand. To the best of our knowledge, our work in this paper is the first one that combines the non-stochastic MAB framework with an ensemble of Transformer networks for query prediction with real-time user feedback adaptation.

\section{Ensemble-based Sequential Query Prediction with Immediate Feedback}
\label{sec:algo}
In this section, we introduce our MAB mechanism based on an ensemble of language model experts, which leverages immediate user feedback for the next query prediction tasks. The corresponding algorithm and regret analysis of our mechanism are reported to provide insights into the capacity of our mechanism to build incrementally candidate query sets for prediction. 

\subsection{Non-stochastic MAB with Transformer-based Experts and Feedback}
Our mechanism, coined \textbf{\expte} for non-stochastic MAB with \textbf{T}ransformer-based \textbf{E}xperts and \textbf{F}eedback, is provided in Algorithm~\ref{alg:qr_sse}. The fundamental feature of \expte is that the set of queries is not fixed and not given beforehand to the algorithm. Specifically, for each user session, \expte builds a candidate query set incrementally, starting with an empty set $\sC_{0}$. Then, \expte starts off the learning process with an ensemble of Transformer-based experts $\sE$, a recommendation threshold $k$, and a learning rate $\eta \in (0,0.5)$. 

At each time step $t>0$ of the learning process, \expte observes the currently executing query $q_t$ selected by the user. Then, using the query context $q_{1}, \ldots, q_{t}$, \expte probes 
each of the Transformer-based experts for the next queries to be predicted and then decides on the corresponding recommendation. These predicted queries are added to the candidate set. The algorithm then estimates the current distribution over the set of these queries in the candidate set, which is the basis for recommending a query by sampling from this distribution.  

\begin{algorithm2e}[tb!]
    \caption{\expte-MAB Query Prediction w.r.t. Transformer-based Experts \& Feedback}
    \label{alg:qr_sse}
    \SetKwInOut{Parameters}{Parameters} \SetKwInOut{Output}{Output}
    \SetKwInOut{Initialization}{Initialization}
    \SetKwFunction{To}{to} \SetKwFunction{score}{Score}
    \SetKwFunction{Threshold}{Threshold} \SetKwFunction{and}{and}
    \Parameters{Set of experts $\sE$, learning rate $\eta \in (0,0.5)$, recommendation threshold $k$}
    \For {step $t= 1, 2,\dots K$} {
        Get current executing query $q_t$ and form the query context $v_t$\; \label{line:two}
        $\sQ_t = \bigcup\limits_{e \in \sE}f(e,v_t,k)$ \tcp*{API call to Transformer-based experts} \label{line:three}
        $\sC_{t} = \sQ_t \cup \sC_{t-1}$\; \label{line:four}
        
        /*Assign and update query weights*/\;
        \If {$(t == 1)$} {
            $w_{i,t} = \frac{\eta}{(1-\eta)\lvert \sC_t \rvert}, \qquad \forall i \in \sC_t$ \; \label{line:five}
            }
        \Else {    
            $w_{i,t} = \frac{\eta}{1-\eta} \frac{\mathds{I}\{i \in \sC_t \setminus \sC_{t-1}\}}{\lvert \sC_t \setminus \sC_{t-1}\rvert}\sum_{i \in \sC_{t-1}} \hat{w}_{i,t} + \hat{w}_{i,t} \mathds{I} \{i \in \sC_{t-1}\}$\; \label{line:six}
        }
        
        Estimation of query probability distribution $p_{i,t} = \frac{(1-\eta) w_{i,t}}{\sum_i w_{i,t}} + \frac{\eta}{\lvert C_t\rvert} \qquad \forall i \in \sC_t$ \; \label{line:seven}
        Sample $I_t$ according to distribution $p_{t}$ \; \label{line:eight}
        Recommend the query $q^{\prime}_{I_t} \in \sC_t$  and capture user reward $r_{I_t,t}$ \; \label{line:nine}
        Calculate the pseudo-rewards:  
        $\hat{r}_{j,t} = 
            \begin{cases} 
                \frac{r_{j,t}}{p_{j,t}}, & \text{if $j = I_t$} \\
                0, & \text{else} 
                \end{cases}
        $\; \label{line:ten}
        Update the query weights based on pseudo-rewards: 
        $\hat{w}_{j,t+1} = 
            \begin{cases}
                w_{j,t}\cdot \exp{(\eta \hat{r}_{j,t})}, & \text{if $j = I_t$} \\
                w_{j,t}, & \text{else}
            \end{cases}
                $\; \label{line:eleven}  
        }
\end{algorithm2e}

In detail, at the time $t$, the algorithm observes the currently executing query $q_t$ (line:\ref{line:two}). The $q_t$ can be either the free text entered by the user or one of the recommended queries in the previous step $t-1$. 
The sequence of queries issued by the user, i.e., the combination of currently executing query $q_t$ and the queries issued till the time step $t-1$, form the current query context, i.e., the sequence $v_t = (q_1, q_2,\dots q_t)$. In line:\ref{line:three}, we call the function $f(e, v_{t}, k)$, which is a standard API function call for a finite set $\mathcal{E}$ of Transformer-based experts $e \in \mathcal{E}$.
The function takes the query context $v_t$, expert instance $e$, and a threshold $k$ as the arguments and returns the predicted queries from the expert $e$. 
For example, if $e$ is an instance of a GPT model in \cite{radford2019language}, the function returns the predicted queries obtained by GPT with current query context $v_t$ as the input. The $k$ argument controls the number of the predicted queries for each expert. It can be either a probability threshold or a count.
In the case of probability threshold, the expert returns the predicted queries whose probability of relevance to the query context is at least $k >0$. In the case of the count, 
this corresponds to a fixed number of predicted queries returned by the expert, e.g., the top-5 predicted queries. In our experiments, we set $k$ as a count threshold. The rationale of a count threshold is that, instead of simply obtaining only the top recommendation, which is not an efficient approach as described above, \expte considers \textit{all} the available query predictions that satisfy the threshold $k$. We carry out this step for all the experts, and the set of the predicted queries $\sQ_t$ is obtained by taking the union of the individual predictions per expert.

In turn, in line:\ref{line:four}, the candidate query set is updated so that the candidate set at step $t$ is obtained by combining different experts' predictions at step $t$ along with the candidate query set at step $t-1$. In line:\ref{line:five} and \ref{line:six}, we define certain weights for the queries in the candidate set. We use a fixed weight for the newly predicted queries (at time $t$), whereas, for the older ones (at times $<t$ ), we adopt the same weight as in the previous rounds.
Specifically, at step $t$, queries that are part of the candidate queries in step $t-1$, we use the same weights as those at the end of step $t-1$. While, for the newly added queries, i.e., those which are not part of the candidate query set at step $t-1$ but added at step $t$, a weight value is assigned factored by $\frac{\eta}{1-\eta}$ and divided by the number of the new queries in the candidate set. The weights will be used for sampling a query from the candidate set for the query prediction. 

Given the updated weights, the algorithm estimates the current probability distribution $p_{t}$ over the predicted queries in the candidate set. At each step, $t$, the algorithm samples an index $I_t$ from the candidate set according to the distribution $p_{t}$ that is proportional to the assigned weights at time $t$ and \textit{recommends} the query ${q^{\prime}_{I_t}}$ to the user (line:\ref{line:seven}). 
The algorithm then observes the reward for the recommended query and calculates the \textit{pseudo} reward (line:\ref{line:ten}) as in the standard expert-based algorithms. The pseudo-reward is an unbiased estimator of the reward. 
Then, depending on the user feedback, the weights are updated at the end of each round to \textit{reflect} that feedback (line:\ref{line:eleven}). 
To update the weights, one can use any commonly employed \textit{potential function} \cite{cesa2006prediction} like polynomial potential function or exponential potential function.
Our algorithm's weights are updated using the exponential weighting scheme due to efficient regret performance \cite{cesa2006prediction}. 

\subsection{\expte Regret Analysis}
The performance of online algorithms is evaluated using regret. Formally, regret is defined as the loss suffered by the algorithm relative to the optimal strategy. After $T$ rounds, the cumulative regret of an online algorithm is defined as: 
\begin{equation}
    R(T) = T - \sum_{t=1}^T r_t.
    \label{eq:regret}
\end{equation}

We prove that the cumulative regret in (\ref{eq:regret}) of the proposed \expte matches with the regret order of the INF algorithm proposed in \citep{audibert2009minimax}. The INF algorithm has the best obtainable regret in the adversarial MAB setting (disregarding the constant factors).

\begin{theorem}
For any $T>0$, $\lvert\sC_{T}\rvert>0$, and learning rate $\eta = \min(0.5, \sqrt{\frac{T+1}{T(2\lvert \sC_T \rvert +1)}})$, the regret of the \expte algorithm is $\mathcal{O}\left(\sqrt{\frac{\lvert \sC_T \rvert}{T}}\right)$. 
\end{theorem}

\begin{proof}
Our proof is based on the assumption that at any time $t$, the candidate set contains the optimal query. The proof extends the proof of \expt to handle the dynamic nature of the candidate set. At each step $t$, the number of queries to be recommended may change.
We denote the set of candidate queries available at time step $t$ as $\sC_t$.
For convenience in analysis, we assume that $\lvert \emptyset \rvert = 1$.

Let us define the variable $W_t$ as the sum of the weights of the queries in the candidate set $\sC_t$ at the end of the $t$-th round, i.e., $ W_t = \sum_{i \in \sC_t} w_{i,t}$. Firstly, we derive a lower bound such that:

\begin{align}
\ln{\frac{W_{T+1}}{W_1}} &= \ln{\sum_{i \in C_{T+1}} w_{i,T+1}} - \ln{W_1} \label{eq:no1} \text{. Then, }\\
\ln{\frac{W_{T+1}}{W_1}} &\geq \ln{w_{j,{T+1}} } - \ln{\frac{\eta}{1-\eta}} \qquad \text{for any } j \in \sC_{T+1}  \label{eq:no2} \\
&= \eta \sum_{t=1}^T\hat{r}_{j,t} - \ln{\frac{\eta}{1-\eta}} \label{eq:no3}. \qquad \text{ Thus,}\\
\ln{\frac{W_{T+1}}{W_1}} &\geq \eta \sum_{t=1}^T\hat{r}_{j,t}  - \ln{\frac{\eta}{1-\eta}} \label{eq:no4}.
\end{align}

It should be noted that (\ref{eq:no2}) holds for any query in the candidate set (and not just for the query that is recommended). To emphasize this, we use the index $j$.
Secondly, we bound the weights to derive a suitable upper bound for $t < T$, i.e.,
\begin{align}
\ln{\frac{W_{t+1}}{W_t}} &= \ln{\left(\sum_{i \in \sC_{t+1}} \frac{w_{i,t+1}}{W_t}\right)} \\
&= \ln{\left(\sum_{i \in \sC_t} \frac{\hat{w}_{i,t+1}}{W_t} + \sum_{i \in \sC_{t+1} \setminus \sC_t}\frac{w_{i,t+1}}{W_t} \right)}  \\
&= \ln{\left(\sum_{i \in C_{t}} \frac{\hat{w}_{i,t+1}}{W_t} + \frac{\eta}{1-\eta} \right)} \\
& = \ln{\left(\sum_{i \in \sC_t}  \frac{w_{i,t}\exp{(\eta\hat{r}_{I_t,t})}}{W_t} + \frac{\eta}{1-\eta} \right)}
\end{align}
By using $\lvert \sC_t\rvert \geq 1$ and $w_{i,t} = \frac{W_t}{1-\eta}\Big(p_{i,t} - \frac{\eta}{\lvert \sC_t \rvert}\Big)$, and approximating $e^{x} \leq 1+x+x^2 \, \text{ for }\, x < 1.79$, we obtain that:

\begin{align}
\ln{\frac{W_{t+1}}{W_t}} &\leq \ln{\left(\frac{\eta}{1-\eta} + \frac{1}{1-\eta}\sum_{i \in\sC_t}\left(p_{i,t} -\frac{\eta}{\lvert \sC_t \rvert}\right)\left( 1+\eta\hat{r}_{I_t,t} + \eta^2\hat{r}^2_{I_t,t}\right)\right)}  \label{eq:no5}\\
&\leq \ln{\left(\frac{\eta}{1-\eta} + \frac{1}{1-\eta}\sum_{i \in\sC_t}\left(p_{i,t}+\eta p_{i,t}\hat{r}_{I_t,t} + \eta^2 p_{i,t} \hat{r}^2_{I_t,t} - \frac{\eta}{\lvert \sC_t \rvert}\right)\right)}  \label{eq:no6}\\
&= \ln{\left(\frac{\eta}{1-\eta} - \frac{1}{1-\eta}\sum_{i \in \sC_t} \frac{\eta}{\lvert \sC_t \rvert}+ \frac{1}{1-\eta}\left(\sum_{i \in\sC_t} \left(p_{i,t} + \eta p_{i,t}\hat{r}_{I_t,t} + \eta^2 p_{i,t} \hat{r}^2_{I_t,t} \right)\right)\right)}  \label{eq:no7} \\
&= \ln{\left(\frac{1}{1-\eta} + \frac{1}{1-\eta}\sum_{i \in\sC_t} \left( \eta p_{i,t}\hat{r}_{I_t,t} + \eta^2 p_{i,t} \hat{r}^2_{I_t,t} \right)\right)}  \label{eq:no8} \\
&\leq \ln{ \left( 2 + \frac{1}{1-\eta}\sum_{i \in\sC_t} \left( \eta p_{i,t}\hat{r}_{I_t,t} + \eta^2 p_{i,t} \hat{r}^2_{I_t,t} \right)\right)}  \label{eq:no9} \\
&\leq 1 + \frac{\eta}{1-\eta} \sum_{i \in C_t} p_{i,t} \hat{r}_{I_t,t} + \frac{\eta^2}{1-\eta} \sum_{i \in \sC_t} p_{i,t}\hat{r}_{I_t,t}^2  \label{eq:no10} \\
&= 1 + \frac{\eta}{1-\eta} r_{I_t,t} + \frac{\eta^2}{1-\eta} \sum_{i\in \sC_t} \hat{r}_{I_t,t}\mathds{1}\{i == I_t\} \label{eq:no11}
\end{align}

In \eqref{eq:no5}, we expanded the equation by multiplying respective terms. In \eqref{eq:no6}, we used the fact that $\eta$, and $\hat{r}_{I_t,t}$ are non-negative values. In \eqref{eq:no8}, we used the fact that $\sum_{i \in\sC_t}  p_{i,t} =1$ and $\sum_{i \in\sC_t} \frac{\eta}{\lvert \sC_t \rvert} = \eta$. In \eqref{eq:no10}, we used the inequality $\ln{(1+x)} \leq x \, \text{ for }\, x > 0$ and in \eqref{eq:no10} we assume that rewards are bounded in $(0,1)$. $\mathds{1}\{i == I_t\}$ in \eqref{eq:no11} denotes the indicator variable. By summing over $t$, we obtain: 

\begin{align}
\ln{\frac{W_{T+1}}{W_1}} & \leq T + \frac{\eta}{1-\eta} G_T + \frac{\eta^2}{1-\eta}\sum_{t=1}^T\sum_{i \in \sC_t} \hat{r}_{I_t,t}\mathds{1}\{i == I_t\} \label{eq:no12}\\
&\leq T + \frac{\eta}{1-\eta} G_T + \frac{\eta^2}{1-\eta}\lvert\sC_T\rvert \sum_{t=1}^T \hat{r}_{I_t,t}. \label{eq:no14}
\end{align}

In \eqref{eq:no12}, $G_{T}$ refers to the sum of the rewards, i.e., $G_{T} = \sum_{t=1}^T r_{I_t,t}$. We then combine the lower bound in (\ref{eq:no4}) and upper bound in (\ref{eq:no14}), thus, obtaining,

\begin{align}
\eta \sum_{t=1}^T\hat{r}_{j,t}  - \ln{\frac{\eta}{1-\eta}} & \leq \frac{\eta}{1-\eta} G_T + \frac{\eta^2}{1-\eta}\lvert\sC_T\rvert \sum_{t=1}^T \hat{r}_{I_t,t} + T  \label{eq:no15} \qquad \text{or,}\\
\frac{\eta}{1-\eta} G_T + \ln{\frac{\eta}{1-\eta}} & \geq   \eta \sum_{t=1}^T\hat{r}_{j,t} - \frac{\eta^2}{1-\eta}\lvert\sC_T\rvert \sum_{t=1}^T \hat{r}_{I_t,t} - T \label{eq:no16} \\
\frac{\eta}{1-\eta} G_T + 1 & \geq   \eta \sum_{t=1}^T\hat{r}_{j,t} - \frac{\eta^2}{1-\eta}\lvert\sC_T\rvert \sum_{t=1}^T \hat{r}_{I_t,t} - T \label{eq:no17} \\
\end{align}
In \eqref{eq:no17}, we used the fact that ,for any plausible values of $\eta$, $\ln{\frac{\eta}{1-\eta}} \leq 1$. Taking the expectation in both sides, we get 
\begin{equation}
\begin{split}
\frac{\eta}{1-\eta} \mathbb{E}[G_T] +1 & \geq   \eta \sum_{t=1}^T r_{j,t} - \frac{\eta^2}{1-\eta}\lvert\sC_T\rvert \sum_{t=1}^T r_{I_t,t} -T. \\
\end{split}
\end{equation}
Based on the fact that the inequality holds for any $j \in \sC_t$ and the assumption that the optimal query always exists in the candidate set to obtain, we have that,

\begin{equation}
\begin{split}
\frac{\eta}{1-\eta} \mathbb{E}[G_T] +1 & \geq   \eta G^\ast - \frac{\eta^2}{1-\eta}\lvert\sC_T\rvert G^\ast -T \label{eq:no18}, \\
\end{split}
\end{equation}
where in \eqref{eq:no18}, $G^\ast = \max_j \sum_{t=1}^T r_{j,t}$ is the cumulative reward from recommending the single globally best query.

As the rewards are bounded in (0,1), the maximum cumulative reward is therefore bounded by $T$, i.e., $G^\ast \leq T$.  We can then re-write \eqref{eq:no12} and obtain,

\begin{equation}
    \begin{split}
    \eta G^\ast  \leq 1 + \frac{\eta}{1-\eta}\mathbb{E}[G_T] +  \frac{\eta^2}{1-\eta}\lvert \sC_T \rvert T + T.\\ 
    \end{split}
\end{equation}
By converting the \textit{gain} into \textit{loss}, i.e., $L_T = T - G_T$ and $L^\star = T - G^\star$, and using the facts that  $\frac{1}{1-\eta} \leq 2$ and $\mathbb{E}[G_T] \leq T$, we obtain:
\begin{equation}
\begin{split}
\mathbb{E}[L_T] - L^\star &\leq \frac{T+1}{\eta} + 2\eta\lvert \sC_T \rvert T.
\end{split}
\end{equation}
By taking a learning rate $\eta = \sqrt{\frac{T+1}{2T\lvert \sC_T \rvert}}$, we obtain that the per-round regret: 
\begin{eqnarray}
\frac{L_T - L^\star}{T} \leq \mathcal{O}\left(\sqrt{\frac{\lvert \sC_T \rvert}{T}}\right).
\label{eq:endofproof}
\end{eqnarray}
\end{proof}

We can observe that the regret of \expte matches the best obtainable regret in the adversarial bandit setting (disregarding the constant factors) in \citep{audibert2009minimax}, and it is better than the regret obtained using the baseline \expt algorithm \citep{auer2002nonstochastic}, which the later assumes that the candidate set is known in advance. 

\begin{remark}
The \expt \citep{auer2002nonstochastic} algorithm with the known candidate set and known cardinality $m$ achieves a regret of the order $O\left(\sqrt{\frac{m \ln{m}}{T}}\right)$, which is $\ln{m}$ higher than the regret obtained by our \expte algorithm with $m = \lvert \sC_T \rvert$.
\end{remark}

\section{Experimental Evaluation}
\label{sec:exp}
In this section, we report on the dataset, baseline approaches and models under comparison, and the performance metrics adopted in our experimental study. 
Furthermore, we provide an in-depth analysis of the results of different query prediction and recommendation mechanisms using the Transformer-based models. For reproducibility, our source code is publicly available at:~\textbf{\url{https://github.com/shampp/tef}}.

The purpose of this experimental study is threefold: 
We empirically show that (i) Transformer-based models trained with CLM objective perform better than Transformer-based models trained using Masked Language Modeling (MLM) objective; (ii) adversarial MAB strategies by combining the query predictions from an ensemble of experts outperform strategies based on recommendations using a single CLM-based Transformer model as an expert, and (iii) Transformer-based models trained using limited historical query logs slightly underperform publicly available pre-trained Transformer models trained on huge corpus and fine-tuned for query recommendations.

\subsection{Datasets \& Protocol}
Our experiments are conducted on a real large-scale query log from an online literature discovery service.
The dataset is used in the past \citep{parambath2021maxutility} to test stochastic MAB algorithms for query recommendation.
The online literature discovery service lets the user/analyst start the knowledge-gathering session by entering a free-text query. In each subsequent step, the user either chooses from the recommended queries or inputs another free-text new query. The logs contained 4.5 million queries grouped into 547,740 user sessions. We followed the same data preparation techniques as in the paper \cite{parambath2021maxutility}. 
As the pre-processing step, we removed user sessions with less than four queries.
We also removed sessions containing non-English queries. 
The statistics of the final data used in the experiment are given in \autoref{tab:data_stat}. We refer the readers to \cite{parambath2021maxutility} for further details about the data.

\begin{table}
        \centering
        \caption{Experimental Dataset Statistics.}
        \label{tab:data_stat}
        \begin{tabular}{@{}ll@{}}
            \toprule
            Description & COUNT \\
            \midrule
            \# of queries (original) &  4,687,947\\
            \# queries (pre-processed) & 1,120,461\\
            \# user sessions (\texttt{"})& 159,237 \\
            \# avg queries/session (\texttt{"}) & 7 \\
            \bottomrule
        \end{tabular}
\end{table}

In our experiments, at each round, we select user sessions one by one and recommend the predicted queries in each session. The process is carried out for $T=500$ rounds, where $T$ is the horizon. In each user session, we treat the first query as the initial query. We then recommend predicted queries sequentially. In each round, we treat all the executed queries in the selected user session as the query context $v_{t}$. We then feed the query context to get recommendations from the ensemble of experts. For user session with less than 500 queries, we reset the initial query and the query context to restart the on-line learning process.

\subsection{Performance Metrics}
The performance of online algorithms is evaluated using the regret defined in (\ref{eq:regret}). 
In the predicted query recommendation, we adopt binary rewards, such that the query recommendation algorithm receives a reward $r_{t}=1$ if the recommended query is clicked (accepted) by the analyst at step $t$; $r_{t}=0$ otherwise. In our offline experiments, we mimic the click/accept behaviour using the metric BLEU. BLEU is a well-established metric widely used in evaluating the performance of query recommendation tasks \cite{mustar21_tis,ahmad2019context,ren2018conversational,mustar2020using}.
BLEU was originally proposed to measure the correctness of machine translation tasks \citep{papineni2002bleu} and later adopted for tasks like query recommendation and prediction.
Specifically, we adopt the BLUE metric to determine the reward value $r_{t}$. BLEU is defined as the ratio of the generated words in the queries that are present in the predicted query \textit{and} the observed (actual query issued by the user) query. In our experiments, to calculate the regret, we assign rewards based on the value of the BLEU metric. We assign $r_{t}=1$, if the predicted query and observed query share more than half of the words, and a reward $r_{t}=0$, otherwise. 

\begin{figure*}[t!]
    \begin{minipage}{\textwidth}
        \centering
        \begin{tabular}{@{}c@{}}
            \includegraphics[width=\textwidth]{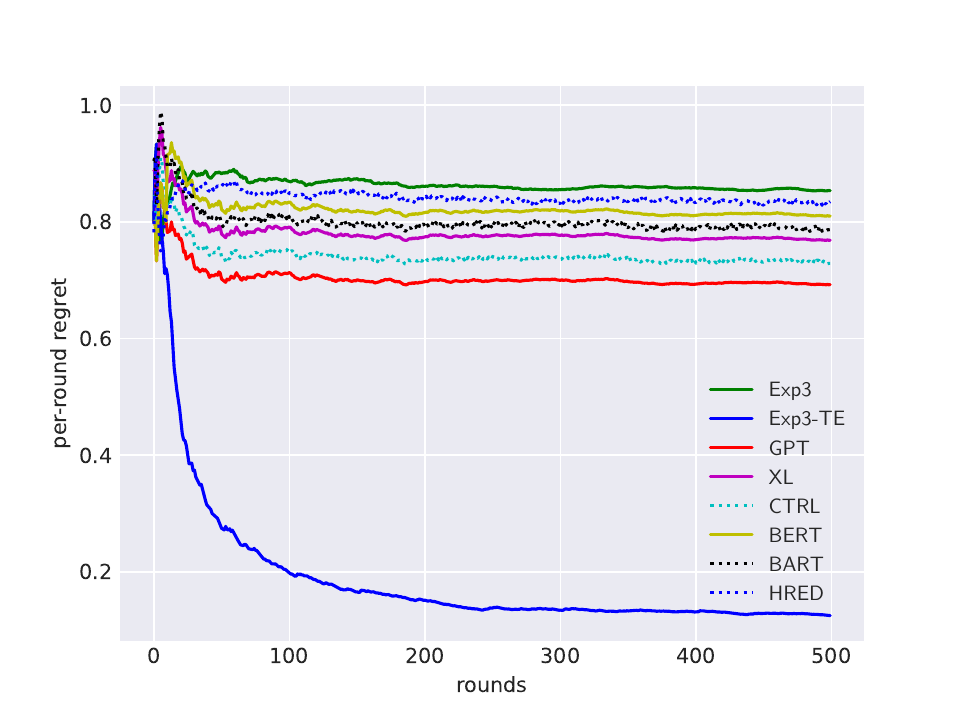}
        \end{tabular}
        \caption{Per-round regret of different comparison algorithms and baselines using \textit{pre-trained} models.}
        \label{fig:pretrained_plot}
    \end{minipage}
\end{figure*}

\begin{figure*}[t!]
    \begin{minipage}{\textwidth}
        \centering
        \begin{tabular}{@{}c@{}}
            \includegraphics[width=\textwidth]{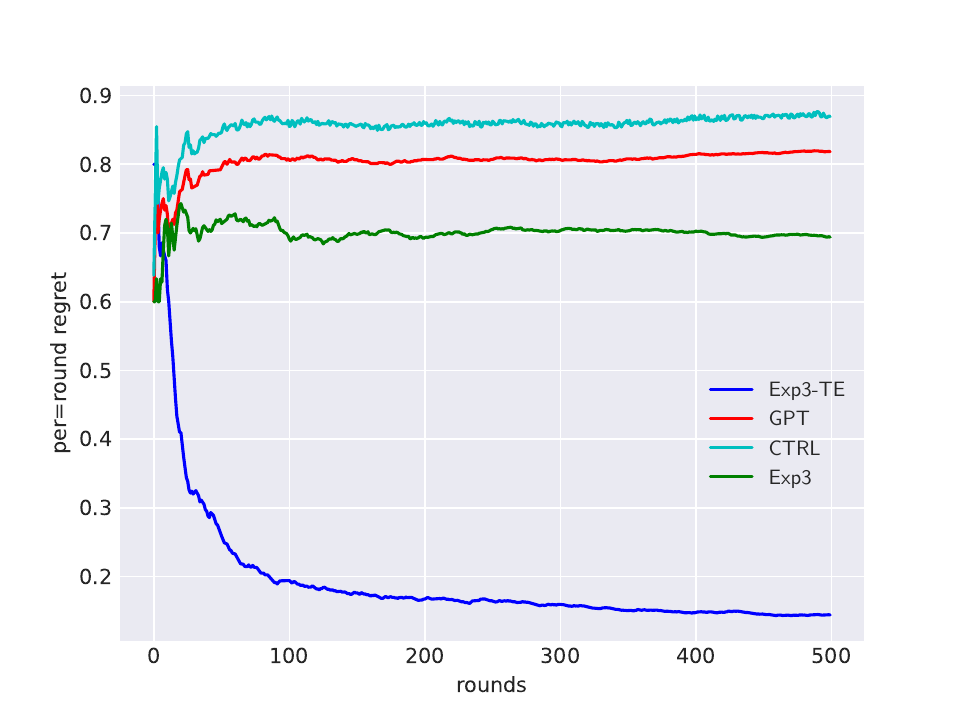}
        \end{tabular}
        \caption{Per-round regret of different comparison algorithms and baselines using models learned \textit{from scratch}.}
        \label{fig:scratch_plot}
    \end{minipage}
\end{figure*}

\subsection{Baseline Approaches \& Models under Comparison}
We conducted a comprehensive experimental evaluation of different models and algorithms with different Transformer-based models. In \citet{mustar21_tis}, authors compared different state-of-the-art query recommendation models, including RNN models, and established that the Transformer model outperforms other models. In our study, we employed three CLM-based models and two MLM-based Transformer models. We used GPT2, CLRT, and Transformer-XL as the CLM models. Moreover, following \citet{mustar2020using,mustar21_tis}, we used BERT and BART as the MLM models. For our proposed algorithm \expte, we used the above-mentioned five Transformer models as experts, i.e., \expte makes use of predictions from an ensemble of five experts: GPT2, CLRT, Transformer-XL, BERT and BART.

We experimented with two different setups for the above-discussed Transformer models.
In the first setup, we used the publicly available pre-trained models from HuggingFace\footnote{\url{https://huggingface.co/models}}.
We fine-tuned the pre-trained models using our training data and used them as the experts for \expte. In the second setup, we trained each of the Transformer models discussed above from scratch using the query logs as the training data. We used the default hyperparameter values for the deep neural networks in both setups. The most important sets of hyperparameter values are given in Table:\ref{tab:train_param}.

\begin{table}
        \centering
        \caption{Training Parameters of the Models under Comparison.}
        \label{tab:train_param}
        \begin{tabular}{@{}lccccc@{}}
            \toprule
                     & GPT2 & CTRL & Trans-XL & BERT & BART\\
            \midrule
            embedding dimensionality  & 1024 & 1280 & 1024 & 768 & 768 \\
            \# layers                 & 24   & 48   &  18  & 12  & 6 \\
            \#  attention heads       & 16   & 16   &  16  & 12  & 12 \\
            dropout probability       & 0.1  & 0.1  &  0.1 & 0.1 & 0.1 \\
            maximum sequence length   & 1024 & 50000 & -   & 512 & 1024\\
            \bottomrule
        \end{tabular}
\end{table}

The comparison models against \expte in our experiments are as follows:
\begin{itemize}
    \item \textbf{GPT2} is a popular Transformer-based language model trained with the CLM objective, i.e., predict the next word given all of the previous words within some text \citep{radford2019language}.
It is, therefore, powerful at predicting the next token in a sequence.
The original GPT2 model has 1.5 billion parameters and is trained on a dataset of 8 million diverse web pages containing English text. The GPT model used in our experiments is available from the HuggingFace interface as: \textit{gpt2-medium}.
\item The \textbf{CTRL} model was proposed in \cite{keskarCTRL2019}. Similar to GPT2, it is a unidirectional Transformer network. CTRL makes use of control codes derived from naturally co-occurring structures in the raw text to generate coherent text.
CTRL is trained over a huge amount of data from different online services like Amazon reviews, Wikipedia, and Reddit. Our CTRL model is available from HuggingFace interface as: \textit{ctrl}.

\item The \textbf{Transformer-XL} model was introduced in \cite{dai2019transformer}. It is a model with relative positioning (sinusoidal) embeddings that can reuse previously computed hidden states to attend to a longer context. Hence, it has the potential to learn longer-term dependency. The Transformer-XL model is trained on WikiText-103 dataset and is available as: \textit{transfo-xl-wt103} in the HuggingFace interface.

\item \textbf{BERT} is a bidirectional (MLM) Transformer model, which is pre-trained using a combination of masked language modelling objective and next sentence prediction on a large corpus comprising the Toronto Book Corpus and Wikipedia \cite{devlin2019bert}.
We used the same pre-trained model as in \cite{mustar2020using}, which is available as: \textit{bert-base-uncased} in the HuggingFace interface.

\item \textbf{BART} uses a standard seq2seq/machine translation architecture with a bidirectional encoder (like BERT) and a left-to-right decoder (like GPT) \citep{lewis2020bart}. The pre-training task involves randomly shuffling the order of the original sentences and a novel in-filling scheme, where spans of text are replaced with a single mask token. The BART model used in our experiments is trained over the CNN/DM dataset and is available as \textit{facebook/bart-base} in the HuggingFace interface.

\item The \textbf{HRED} is a hierarchical recurrent encoder-decoder model introduced in \citet{sordoni2015hierarchical}. Given a user session and any query in that session, HRED encodes the information seen up to that query, i.e., all the previous queries in the session, and then tries to predict the next query, just like in typical language models. But HRED makes use of hierarchical encoder-decoder architecture. The process is iterated throughout all the queries in the session. We used the original implementation provided by the authors \footnote{https://github.com/sordonia/hred-qs}.

\item \textbf{\expt} \citet{auer2002nonstochastic} is the baseline MAB model for \expte. Like in our case, \expt makes use of the immediate user feedback to estimate the sampling distribution to choose the next query to be recommended. The original \expt algorithm works only with a fixed set of arms. Hence, it cannot be used directly in sequential query prediction using an ensemble of Transformer experts. To circumvent this issue, we take at most fifty recommendations from the experts for every initial query and fix this as a static candidate arm set. We chose fifty recommendations, as they can include all the relevant and yet diverse queries pertaining to the initial query. 
\end{itemize}

\begin{remark}
We excluded the following two baseline models from our experimental evaluation for the following reasons. We excluded the continuously re-trained Transformer model.
As a baseline, we could try to retrain or fine-tune any Transformers after accumulating a fixed number of user responses. However, the Transformer models have a large number of parameters which makes constant retraining and fine-tuning significantly inefficient in space and time and contributing to a higher carbon footprint.
Furthermore, it is impractical in knowledge-gathering tasks involving query recommendation applications, as predictions should be served in seconds. 

In addition, we excluded the EXP4 algorithm proposed in \citet{auer2002nonstochastic}. At each time step of the EXP4 algorithm, it gets a single recommendation from the experts with the goal of finding the best expert (regarding the recommendation) in hindsight. As we pointed out in the revised rationale \& motivation, the top-1 recommendation is not always the best one. That is the motivation for taking top-$k$ recommendations from all experts to build the candidate set sequentially. To use EXP4 in our settings, we may have to consider the top-$k$ recommendations from each expert. However, the current literature lacks how one can calculate the rewards for the $k-1$ non-chosen queries from the chosen expert. It is something we are considering as a future work in our agenda. Additionally, in the rationale and motivation section, we exemplified the suboptimal performance of the strategy of recommending from a single expert.

Additionally, EXP4 algorithm is similar to the Hedge algorithm \citep{freund1997decision} and aims to find the best-performing expert from the given set of experts by adapting to the immediate feedback. 
Consequently, EXP4 cannot perform better than the best-performing expert. Since all the individual experts perform poorer than our proposed algorithm, as evidenced in our experimental study (see Figures:~\ref{fig:pretrained_plot} and \ref{fig:scratch_plot}), we exclude EXP4 algorithm as a baseline. 
\end{remark}


\subsection{Experimental Results \& Analysis}

\begin{figure*}[ht]
    \begin{minipage}{\textwidth}
        \centering
        \begin{tabular}{@{}c@{}}
            \includegraphics[width=\textwidth]{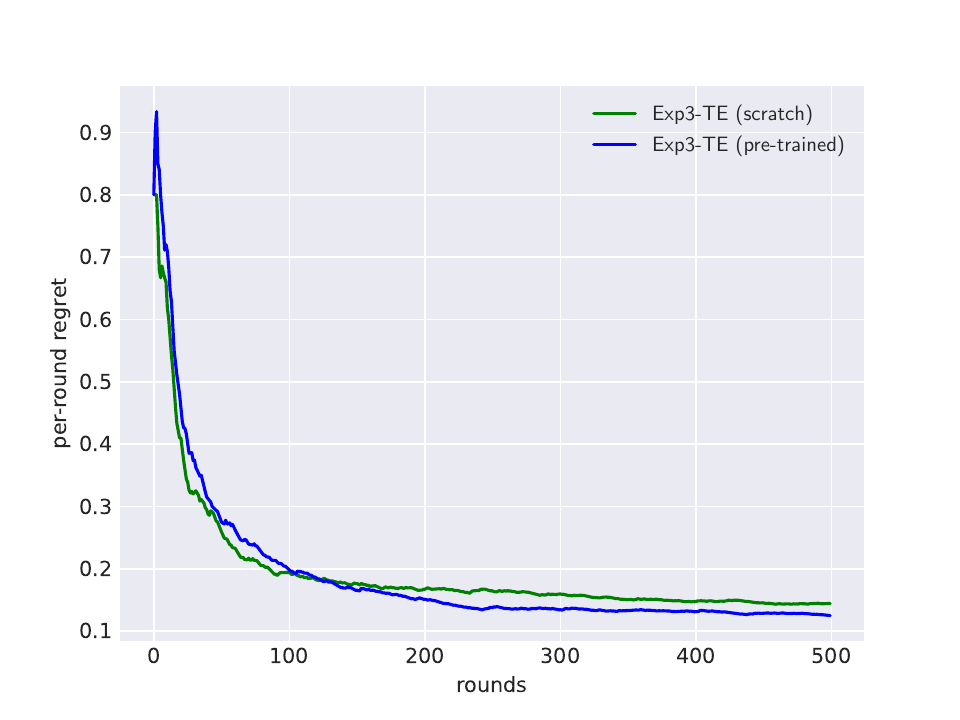}
        \end{tabular}
        \caption{Per-round regret of \expte using both  fine-tuned and trained-from-scratch experts.}
        \label{fig:pre_scratch_plot}
    \end{minipage}
\end{figure*}

\begin{figure*}[ht]
    \begin{minipage}{\textwidth}
        \centering
        \begin{tabular}{@{}c@{}}
            \includegraphics[width=\textwidth]{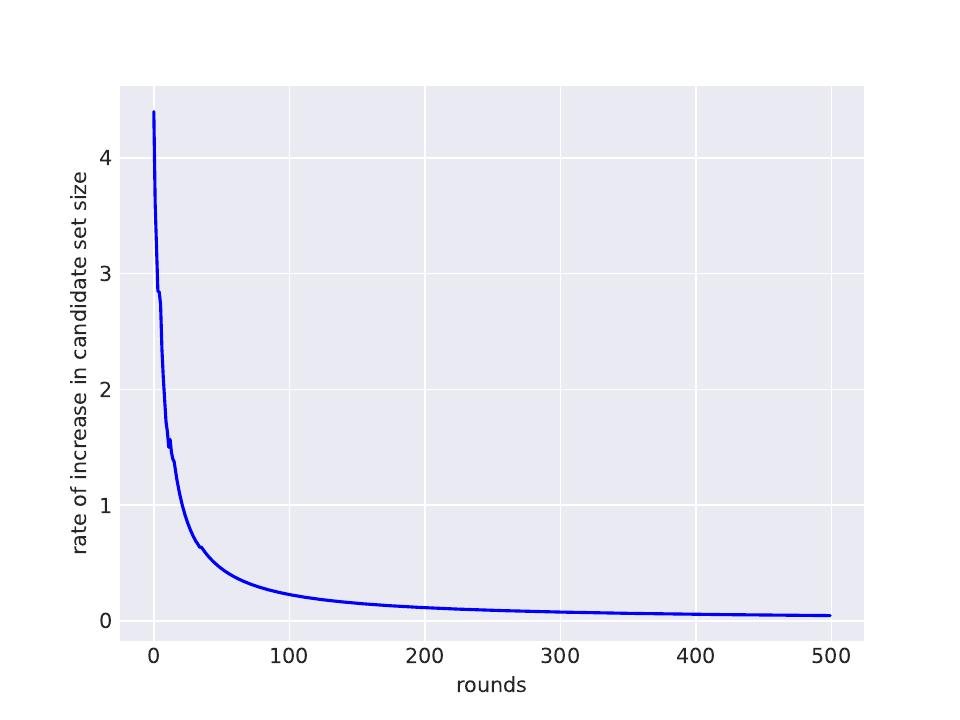}
        \end{tabular}
        \caption{Increase the rate of the candidate set cardinality vs. rounds (using $k=2$ query predictions per expert).}
        \label{fig:cand_rate}
    \end{minipage}
\end{figure*}
We carried out three sets of experiments. 
For all the sets of experiments, we used the top-3 predicted query prediction per expert for \expte, i.e., threshold $k = 3$.
The first set of experiments involved comparing fine-tuned pre-trained CLM models against pre-trained MLM models.
We used our query logs to fine-tune the publicly available models.
These experiments are carried out to show the superiority of the Transformer models trained with CLM objective over the same trained with MLM objective. This set of experiments is inspired by the results of the recent work in \cite{mustar2020using}, where the authors showed that fine-tuning pre-trained MLM-based Transformer models outperform standard RNN-based models. For this set of experiments, we compared the three CLM models against two MLM models used in \cite{mustar2020using}. The results of this experiment are shown in Figure \ref{fig:pretrained_plot}. As it is evident from the plot, CLM-based models (GPT, CTRL, and Transformer-XL) resulted in lower regret than MLM-based models (BERT and BART). Hence, it is advisable to adopt CLM-based transformer models for query prediction as opposed to previous research \citep{mustar2020using}. It is also worth noting that \expt with a fixed candidate set performs poorly compared to other baselines, even though it makes use of user feedback. This clearly demonstrates the efficacy of \expte.

In the second and third classes of experiments, we study the performance of different algorithms using experts learned from scratch.
We compared different CLM-based transformer models trained only on our query logs (no fine-tuning of pre-trained public models).
For this experiment, we used two top-performing CLM models from $1^{st}$ set of the experiment: GPT and CTRL, and trained them from scratch using the query log as the training data.
We used these models learned only from the raw data as the experts in the \expt and \expte algorithms.
The results are given in Figure~\ref{fig:scratch_plot}.
One notable point of this result is that \expt performed better than CLM-based baselines.
This clearly indicates that fine-tuning models trained on vast amounts of data is very important to get good performance.
In both experiments (Figure~\ref{fig:pretrained_plot} and Figure~\ref{fig:scratch_plot}) \expte performed significantly much better than recommending the top result from the experts, irrespective of the fact that the experts are fine-tuned or learned from scratch.
\expte significantly reduces the regret substantially compared to individual expert recommendations by adopting the adversarial recommendation strategy.
This is due to the fact that \expte algorithm is able to recommend the next queries by combining the top predictions from an ensemble of different experts and, thus, increasing the probability of selecting the best recommendation, depending on immediate user feedback and query context. Moreover, \expte is able to exploit not just the top recommendation but the top-$k$ recommendations from different experts. As pointed out earlier, by comparing Figure~\ref{fig:pretrained_plot} and Figure~\ref{fig:scratch_plot}, one can observe that fine-tuning pre-trained models achieve remarkably better regret compared to models learned from limited data.

To directly compare the performance of \expte with and without fine-tuning, we plot the regret of \expte with fine-tuned pre-trained models against \expte with models learned from scratch in Figure~\ref{fig:pre_scratch_plot}.
It can be seen that fine-tuned models give slightly better performance and are preferred over models trained with limited data. 

We also study how \textit{fast} the candidate set is increasing as shown in Figure~\ref{fig:cand_rate}.
After 500 rounds, the average size of the candidate set (overall queries) is 23 (with min and max size values of 8 and 43, respectively). The candidate set remains almost the same after 100 rounds indicating the scalability of our model.

\begin{figure}[ht]
    \begin{minipage}{\textwidth}
        \centering
        \begin{tabular}{@{}c@{}}
            \includegraphics[width=\textwidth]{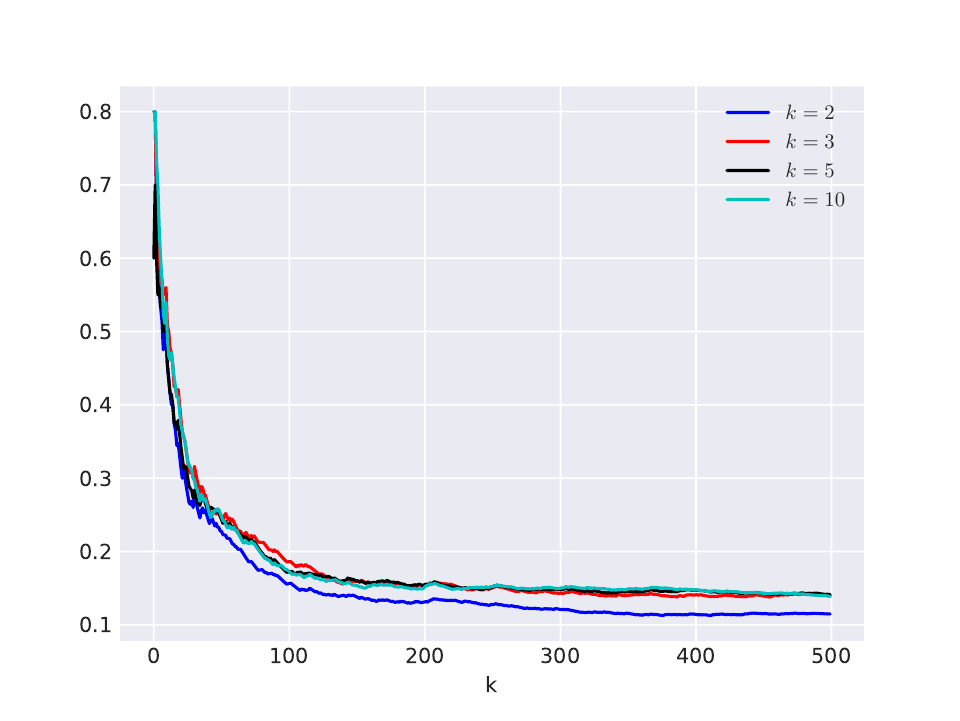}
        \end{tabular}
        \caption{The per-round regret for different values of $k$ in \expte.}
        \label{fig:query_plot}
    \end{minipage}
\end{figure}

Finally, we tested the sensitivity of the number of expert predictions to be included in the candidate set, i.e., how the regret changes as we change the number of predictions used for the query recommendation. In Figure~\ref{fig:query_plot}, we plot the \textit{per-round} regret as the function of the number of the predictions ($k$) used by \expte from the experts.
It is evident that increasing the number of arms (predicted queries) does not significantly improve performance.
It should be noted that there can be duplicate entries from the same expert, and often increasing the number of predicted queries and recommendations does not necessarily increase the size of the candidate set.

\section{Conclusions}
\label{sec:conc}
We introduced the \expte algorithm for query prediction and recommendation tasks in interactive data exploratory analysis processes, like knowledge discovery and information gathering. Our algorithm exploits the adversarial MAB setting with experts' advice, to take into consideration immediate user feedback. Our MAB mechanism augments the Transformer-based language AR models for query prediction and, in turn, recommendations by combining the predictions from different experts and dynamically building a candidate set in each step of the exploration. We make use of immediate user feedback to choose the appropriate recommended query from the candidate set probabilistically. Our experimental results along with a comprehensive comparative assessment using real data showcasing that \expte significantly improves the cumulative regret compared to Transformer-based models and relevant models found in the literature. In our future research agenda, we are investigating non-linear reward structures for query recommendation in the new framework of `deep bandits'. 

\section*{Acknowledgement}
This research is partially funded by the UK `Closed-Loop Data Science for Complex, Computationally-and Data-Intensive Analytics' (\#EP/R018634/1) and the EU Horizon Grant `Integration and Harmonization of Logistics Operations' TRACE (\#101104278). 


\section*{Conflict-of-interest Statement}
The authors have no conflicts of interest to declare.
All co-authors have seen and agree with the contents of the manuscript and there is no financial interest to report. We certify that the submission is original work and is not under review at any other publication.

\bibliography{paper}

\end{document}